\documentclass[letterpaper,10pt]{article}
\usepackage{osameet2}

\usepackage{amsmath,amssymb}
\usepackage{graphicx, caption, subcaption}
\usepackage{hyperref}
\hypersetup{colorlinks=true,citecolor=blue,urlcolor=blue}

\begin{document}

\title{Experimental Demonstration of Data Transmission Based on the Exact Inverse Periodic Nonlinear Fourier Transform}

\author{Jan-Willem Goossens$^{(1,2)}$, Yves Jaou\"en$^{(2)}$ and Hartmut Hafermann$^{(1)}$}
\address{$^1$Mathematical and Algorithmic Sciences Lab, Paris Research Center, Huawei Technologies France SASU\\
$^2$Communications and Electronics Department, Telecom ParisTech, Paris, 75013, France}
\email{jan.willem.goossens@huawei.com}

\begin{abstract}
We design a two-dimensional signal constellation based on the exact periodic inverse nonlinear Fourier transform. Feasibility of continuous transmission with periodic signals is experimentally demonstrated over more than 2000 km.
\end{abstract}


\section{Introduction}

The nonlinear Fourier transform has been studied extensively, both theoretically and experimentally, as a means to cope with fiber Kerr nonlinearity and to ultimately increase data rates beyond that of state of the art linear methods~\cite{Turitsyn2014}. 
The NFT hinges on the fact that despite complex nonlinear interactions between signals in the physical time domain, the spectrum in the nonlinear domain evolves linearly.
Because of the zero boundary conditions and increasing complexity with signal duration of the conventional NFT, burst-transmission is required in practice. The main hurdle in applications is the lack of sufficiently fast NFT algorithms.

The periodic NFT (PNFT) offers a number of advantages over conventional NFT transmission, which holds promise to facilitate practical applications. 
This includes precise control over the signal duration and continuous signaling with cyclic prefix (CP) similar to OFDM instead of the guard band required for conventional NFT transmission. This significantly reduces the processing window and computational complexity associated with the forward transform~\cite{Kamalian2016}.
The underlying algebro-geometric structure~\cite{Belokolos1994} may allow to solve the  multiplexing problem in efficient ways. A fast algorithm for the inverse PNFT however is lacking. An alternative is to precompute nonlinear signals, whose design is a nontrivial problem. 
Ref.~\cite{Kamalian2018} solved the Riemann Hilbert problem to obtain exact periodic solutions of the nonlinear Schr\"odinger equation (NLSE).  Here we design a constellation from finite-gap solutions. Based on PNFT signals stored in lookup tables we for the first time experimentally demonstrate feasibility of continuous PNFT transmission.

\section{Signal design}

Periodic signals can be obtained based on approximate perturbed plane wave solutions, which are constant up to a small periodic perturbation of order $\epsilon$~\cite{Kamalian2016}. The resulting signal is close to amplitude modulation, or, if information is encoded in $\epsilon$, will be hard to detect.
We consider exact periodic solutions of the dimensionless NLSE of the form
\begin{equation}
	\psi(z,t) = U e^{i(k_0z+\Omega_0 t)}\frac{\theta(\frac{1}{2\pi}( k_j z + \Omega_j t + \delta_j^+) | \tau)}{\theta(\frac{1}{2\pi}( k_j z + \Omega_j t + \delta_j^-) | \tau)},\qquad j=1,\,\ldots,\, g,
	\label{eq:thetasolutionSimple}
	\end{equation}
	where $z$ denotes distance along the fiber, $t$ is time and $\theta(u_j|\tau)$ is the $g$-dimensional Riemann Theta function. Moving the phase $\delta^-$ to the numerator eliminates the dependence on the auxiliary spectrum. 
	Given an arbitrary main spectrum with sufficiently distant points $\lambda_{2k}$, $k=1,\ldots g+1$, $\lambda_{2k+1}=\lambda_{2k}^*$, we compute all parameters $\delta_j\equiv \delta^+_j-\delta^-_j$, $U$, $\Omega_j$, $k_j$ and the $g\times g$ period matrix $\tau$ from algebro-geometric loop integrals over the Riemann surface $p^2 = \prod^{2g+2}_{k=1} (\lambda_k-\lambda)$ of genus $g$.

\begin{figure}
     \centering       
                                                                \begin{subfigure}[t]{0.25\textwidth}
                \vspace{2ex}
                                                              \raisebox{-\height}{\includegraphics[width=0.9\textwidth]{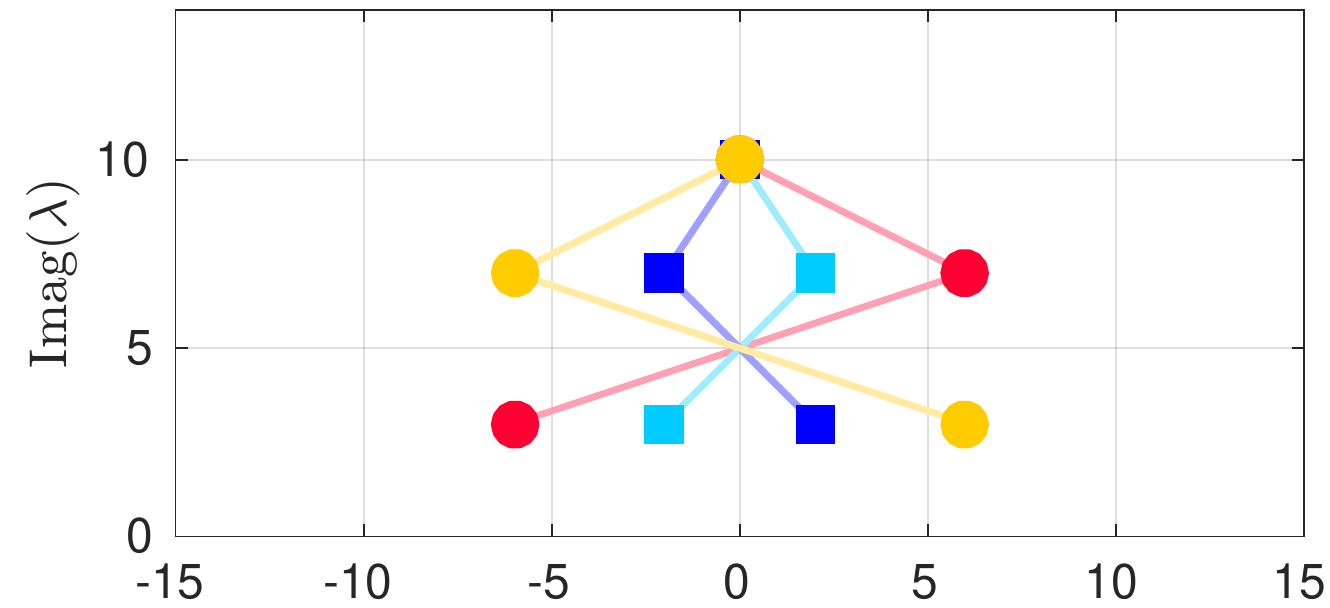}}%
        \vspace{.6ex}
        \raisebox{-\height}{\includegraphics[width=0.9\textwidth]{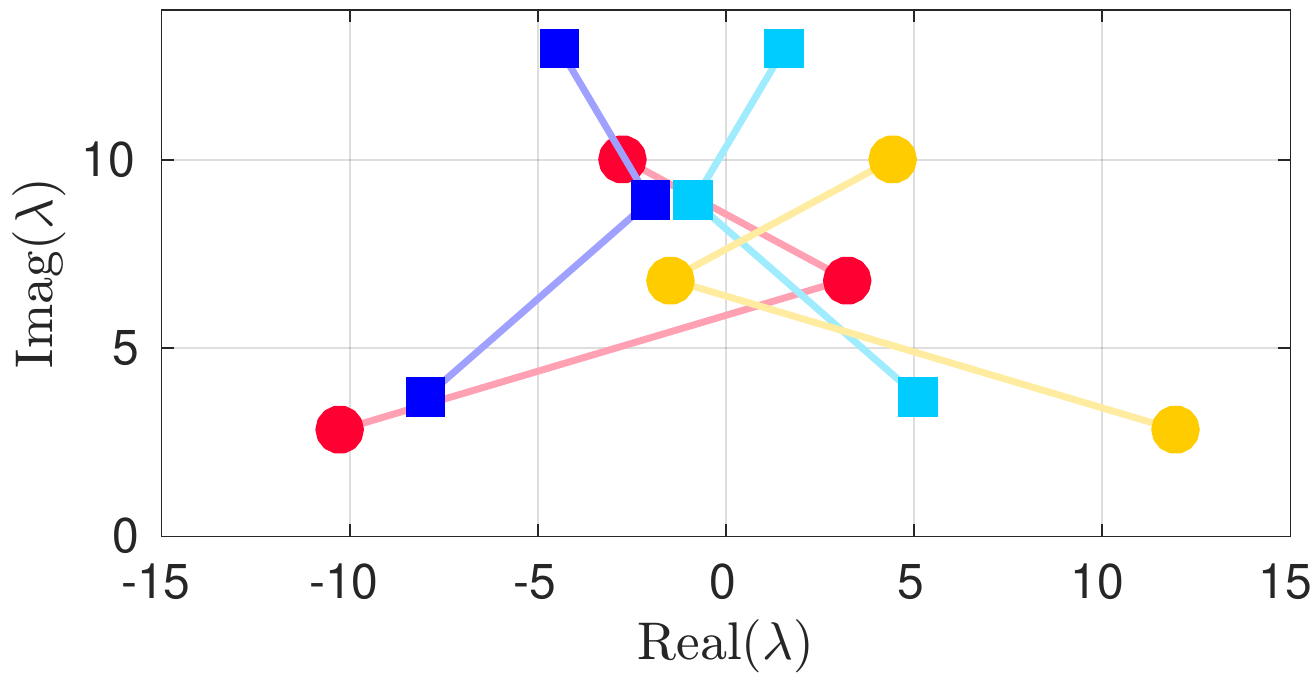}}
    \end{subfigure}
    \hfill
    \begin{subfigure}[t]{0.33\textwidth}
        \raisebox{-\height}{\includegraphics[width=\textwidth]{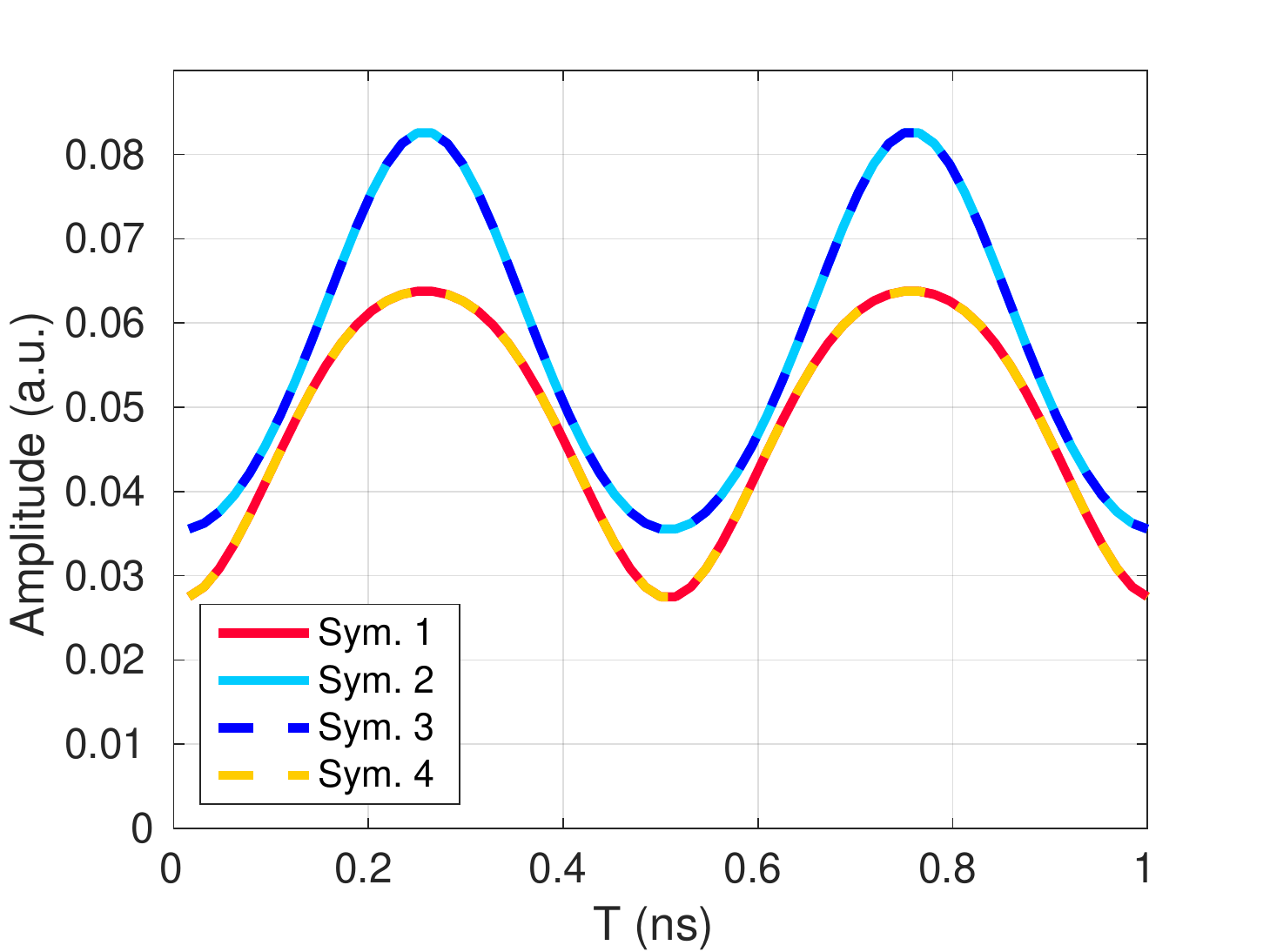}}
    \end{subfigure}
                                \hfill
    \begin{subfigure}[t]{0.33\textwidth}
        \raisebox{-\height}{\includegraphics[width=\textwidth]{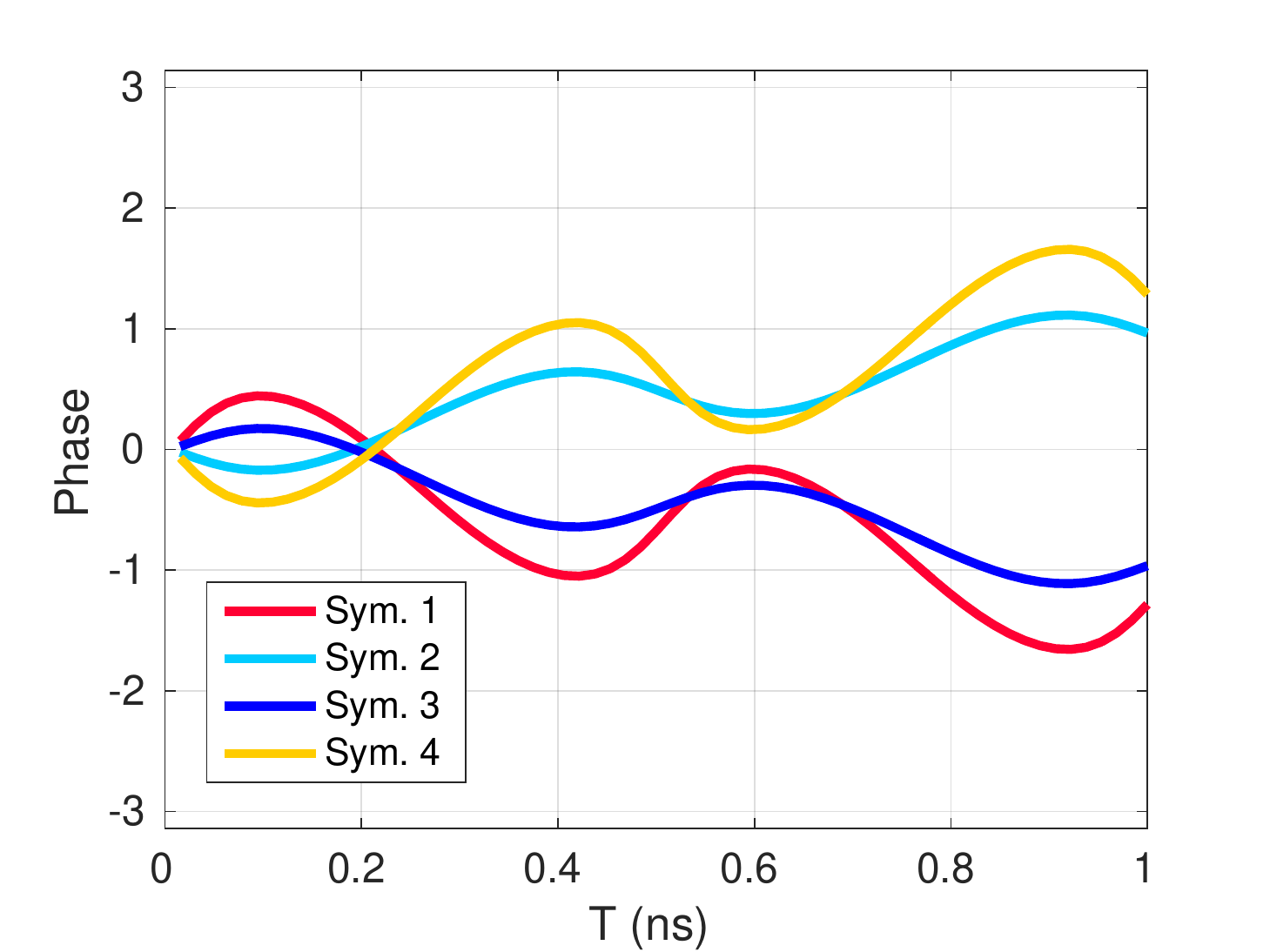}}
    \end{subfigure}
    \caption{From left to right: Signal constellations before (top) and after periodization (bottom), amplitude and phase (before phase matching) of the designed symbols including cyclic prefix.\label{fig:Signals}}
\end{figure}

We design a signal suitable for continuous transmission and finite sampling rate of an arbitrary waveform generator (AWG) and including a CP. 
For a signal whose amplitude evolves non-trivially during propagation, we require signals with genus of at least 2. Obtaining exact periodic signals is cumbersome for spectra of higher genus. For signal recovery via the PNFT it is sufficient to consider quasi-periodic signals for which $\Omega_0$ is not commensurate with $\Omega_j$, $j>0$.
We start with the constellations shown in Fig.~\ref{fig:Signals} (top left). While they are not quasi-periodic in general, one of the frequencies $\Omega_1$, $\Omega_2$ is significantly smaller than the other. By setting it to zero, we obtain a solution which closely approximates a quasi-periodic genus-2 solution with finite period. We obtain symbols with the same physical period of 1 ns including CP through scaling. 
We account for attenuation through an effective nonlinearity parameter based on the transformed-lossless model~\cite{Turitsyn2014}.
Note that one out of three complex degrees of freedom remains unused. In this proof-of-concept study, the constellation consists of four signals with two distinct amplitudes  labeled by a two-bit sequence (signals whose spectrum is mirrored about the imaginary axis are complex conjugate). At the transmitter the signals are sampled at 64 samples/symbol, yielding a data rate of 2Gb/s. The signal occupies a 99\% bandwidth of $B=4.5$ GHz. 
The experimental signal power is fixed to 2.5 dBm and is not a free parameter. 
Fig.~\ref{fig:Signals} shows the resulting constellation and corresponding waveforms. The signals are chained by joining them at their minima to minimize the amplitude mismatch. Unlike solitons, these signals do not decay to zero. Finally, we use the freedom to multiply each signal with a phase factor $\exp(i\phi)$ without changing its main spectrum to make the overall phase continuous.

\section{Experimental setup}

\begin{figure}[htbp]
\centering
\begin{subfigure}{0.2\textwidth}
\includegraphics[width=1.45\columnwidth]{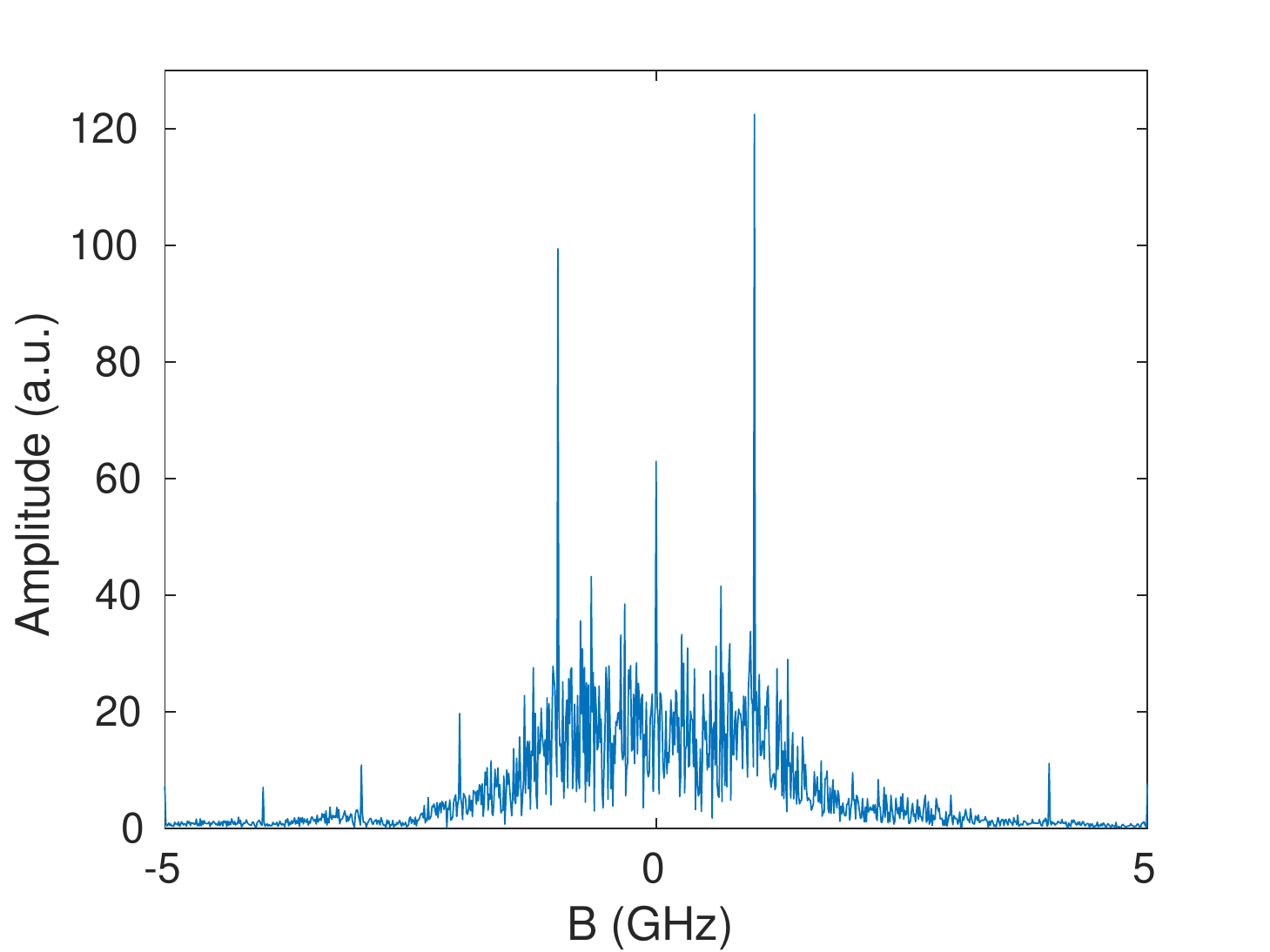}
	\end{subfigure}
	\hspace{4em}
	\begin{subfigure}{0.6\textwidth}
	\includegraphics[width=\columnwidth]{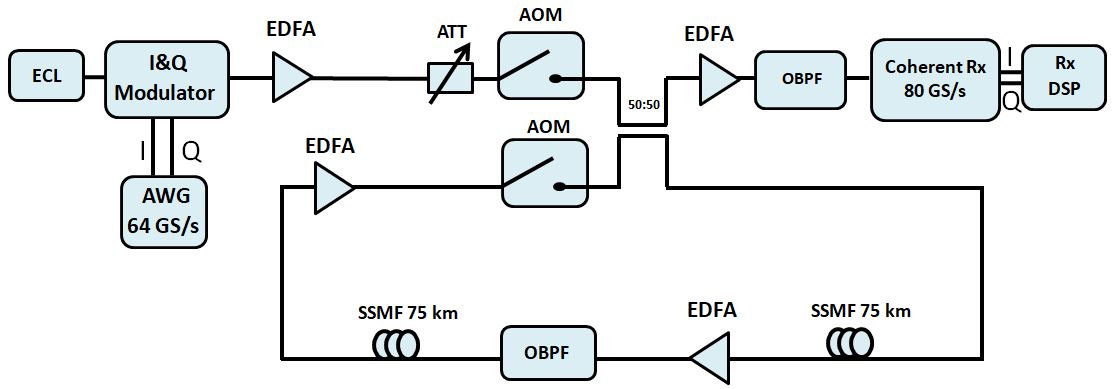}
	\end{subfigure}
\caption{\label{fig:Setup} Transmitted linear frequency spectrum (left) and schematic of the experimental setup (right). The peaks at $\pm$1 GHz correspond to the symbol rate.}
\end{figure}

The experimental setup is shown in Fig.~\ref{fig:Setup}.
At the transmitter (TX) the signal is prefixed with a Schmidl-Cox synchronization symbol and fed into a 64Gs/s AWG 
which drives an I\&Q modulator. The amplitude imperfection of TX and RX is precompensated. 
The signals have slightly different group velocity, which can be tuned by shifting the real part of the nonlinear spectrum. We adjust it manually at the transmitter.
The memory of the AWG limits continuous transmission to 1000 consecutive PNFT symbols.
The optical waveform with spectrum in Fig.~\ref{fig:Setup} (left) is launched into the recirculation loop consisting of two spans of 75 km of standard single mode fiber (SSMF), an Erbium-doped fiber amplifier (EDFA) after each span, and an optical bandpass filter (OBPF) to suppress out of band noise. At the RX the signal is detected in single polarization mode using an oscilloscope with a sampling rate of 80 GS/s.

The RX DSP includes synchronization and compensation of deterministic carrier frequency offset (CFO) caused by the acousto-optic modulators (AOMs). We do not compensate phase noise or stochastic CFO. After signal separation, CP removal, normalization and oversampling to 1024 samples per period we recover the PNFT spectrum using the forward NFT of the FNFT package~\cite{Wahls2018}. We reduce the nonlinear spectrum to the three points with largest imaginary part. Decision on the received symbols $\lambda_k^{\text{recv}}$ is based on the minimal distances between symbols defined as
$\min_{\text{perm}(\lambda)} \sum_{k=1}^{g+1}|\lambda_{k} - \lambda^{\text{recv}}_k|^2$,
where the minimum over all permutations of the three signal points in each signal accounts for the fact that they are indistinguishable.

\section{Experimental results}

\begin{figure}[htbp]
\centering
\begin{subfigure}{0.33\textwidth}
	\includegraphics[width=\columnwidth]{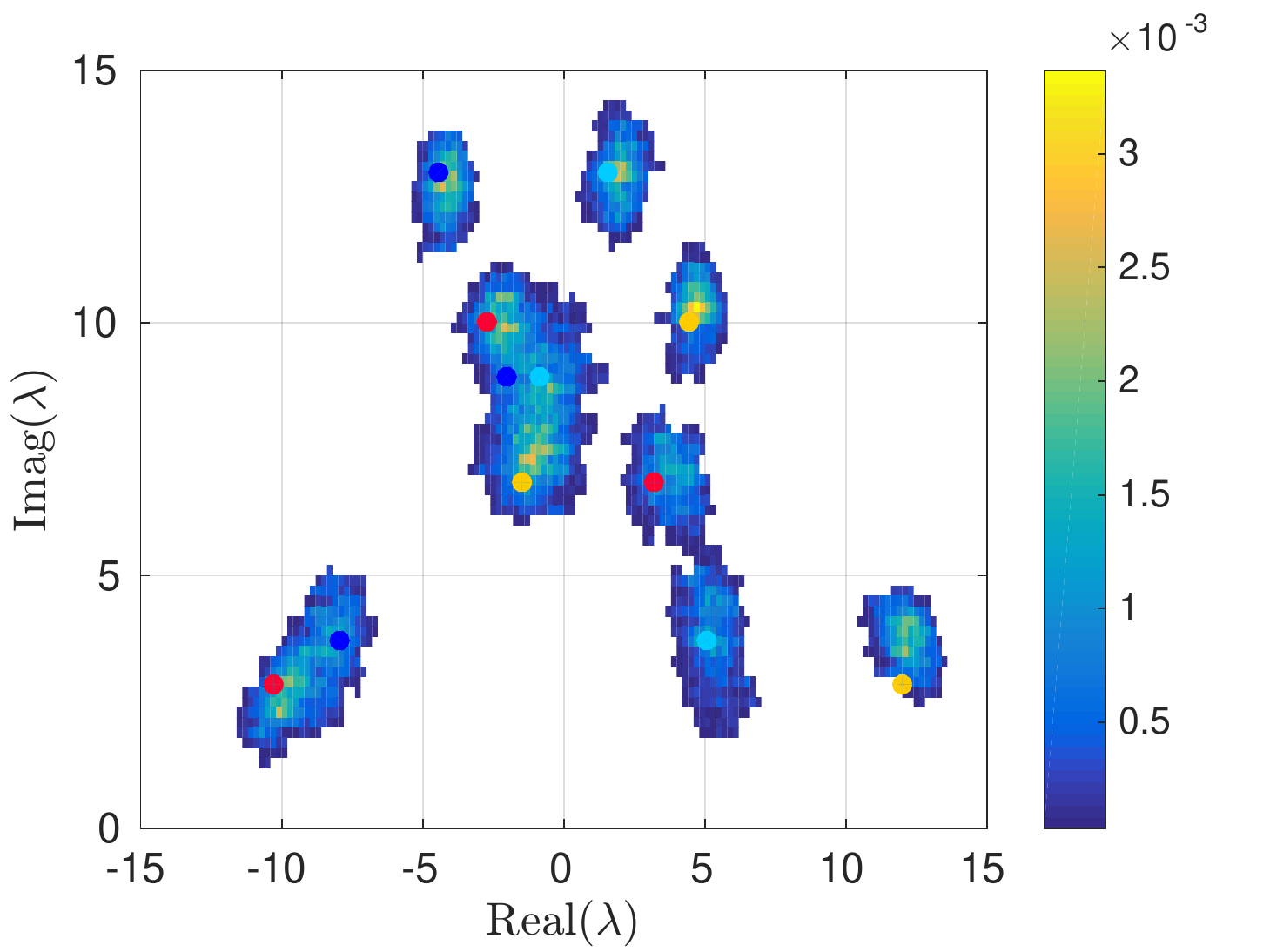}
			\end{subfigure}
	\begin{subfigure}{0.33\textwidth}
	\includegraphics[width=\columnwidth]{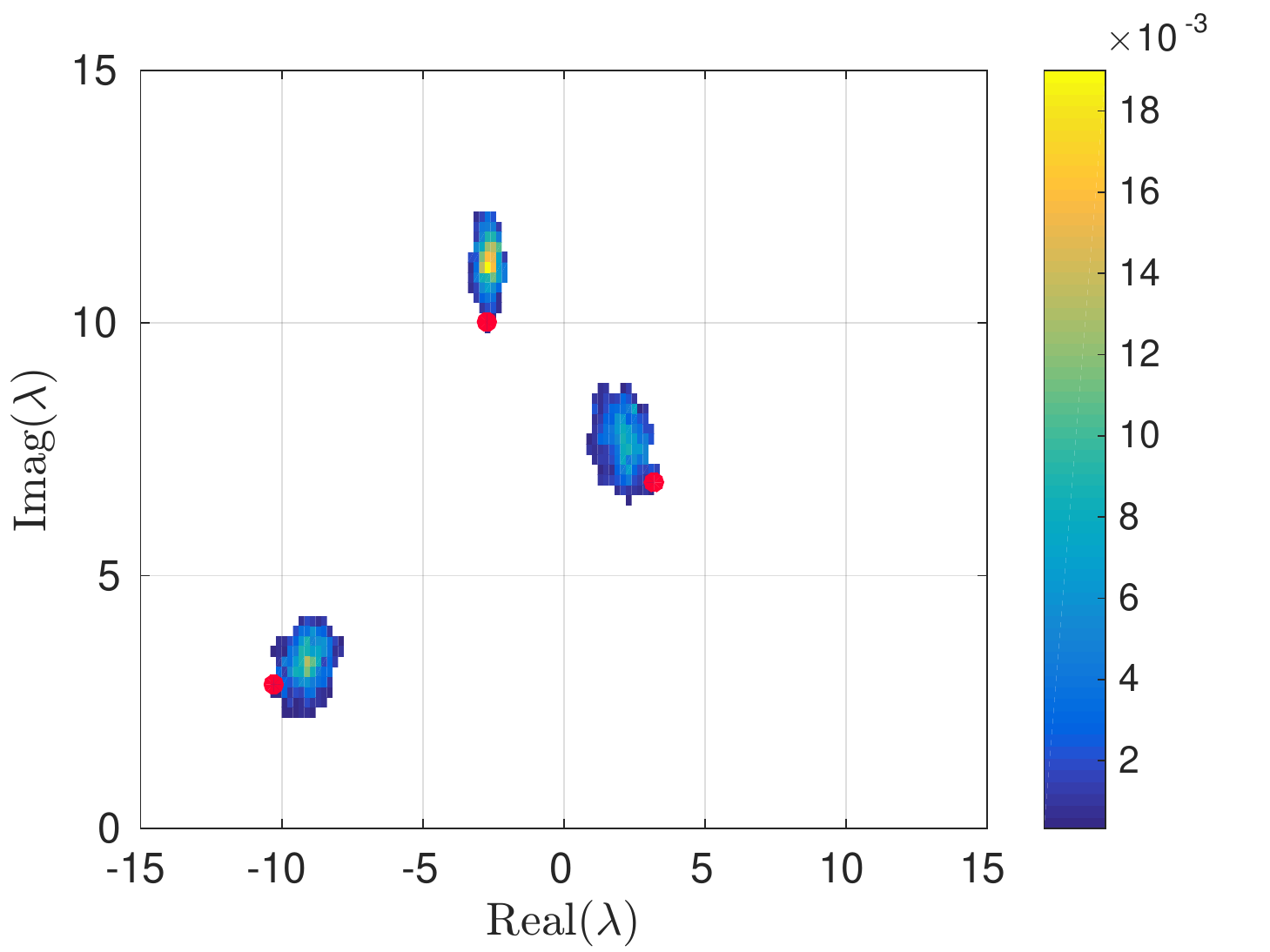}
			\end{subfigure}
	\begin{subfigure}{0.33\textwidth}
	\includegraphics[width=\columnwidth]{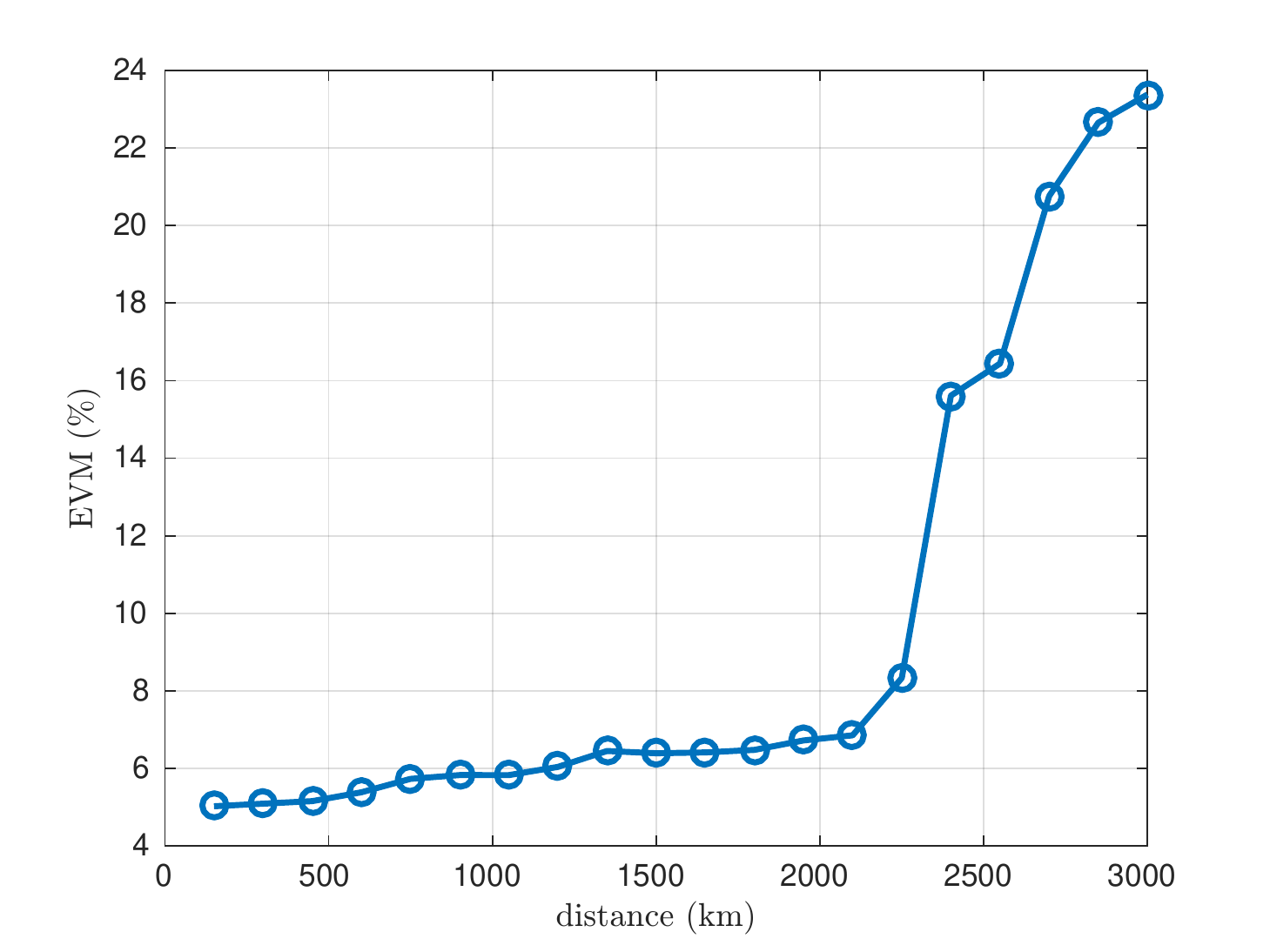}
			\end{subfigure}
\caption{\label{fig:Results} Left: Received constellation after 30 spans (10000 symbols). Colored circles mark the position of the original signal points. Middle: Single signal transmission after 40 spans. Right: EVM as function of transmission distance.}
\end{figure}

Figure~\ref{fig:Results} shows the received constellations (left) after 30 spans (2250 km), corresponding to a BER of 6.4$\cdot 10^{-3}$. The noise distributions clearly depend on the respective signal points. Numerical simulations reveal significant correlations between them. 
 As Fig.~\ref{fig:Results} (right) demonstrates, the error vector magnitude (EVM) 
starts to increase significantly only after more than 30 spans. This is due to inter-symbol interference originating from the mismatch in group velocity and a finite cyclic prefix.
The linear signal broadening expected with a fiber dispersion constant $\beta_2$ is  
$\Delta T=2\pi|\beta_2|L B=1.2$ ns at $L=2250$ km. The deterioration of performance hence happens at distances significantly longer than expected from the length of the cyclic prefix.
Continuous transmission of a single signal (Fig.~\ref{fig:Results} middle) shows that even after 40 spans noise is not the  limiting factor.

\section{Conclusions}
We have designed signals based on the exact inverse PNFT and demonstrated reliable transmission with data recovery in the nonlinear domain over more than 2000 km. 
Precomputed signals have been stored in lookup tables of only 32 complex values.
Our signal design can be generalized to exploit all available degrees of freedom and to significantly improve data rates, spectral efficiency and transmission performance.
The development of an accurate channel model for the PNFT, auxiliary spectrum modulation, and polarization- and wavelength-division multiplexed transmission are left for future work.
\\\\
{\bf Acknowledgements:} We gratefully acknowledge experimental support by W. Gemechu and helpful discussions with S. Chimmalgi, M. Kamalian, P. Prins, S. K. Turitsyn, S. Wahls and M. Yousefi.


\begin{thebibliography}{99}

\bibitem{Turitsyn2014} S. K. Turitsyn, J. E. Prilepsky, S. T. Le, S. Wahls, L. L. Frumin,  M. Kamalian, and S. A. Derevyanko, ``Nonlinear Fourier transform for optical data processing and transmission: advances and perspectives,'' Optica, vol. 4, no. 3, pp. 307--322, (2017).	
\bibitem{Kamalian2016} M. Kamalian, J. E. Prilepsky, S. T. Le, and S. K. Turitsyn, ``Periodic nonlinear Fourier transform for fiber-optic communications, Part I-II,'' Opt. Express, vol. 24, no. 16, pp. 18353--18381, (2016).
\bibitem{Belokolos1994}E. D. Belokolos, A. I. Bobenko, V. Z. Enol'skii, A. R. Its, and V. B. Matveev, ``Algebro-geometric Approach to Nonlinear Integrable Equations,'' Springer-Verlag (1994).
\bibitem{Kamalian2018} M. Kamalian, D. Shepelsky, A. Vasylchenkova, J. E. Prilepsky, and S. K. Turitsyn, ``Communication System Using Periodic Nonlinear Fourier Transform Based on Riemann-Hilbert Problem,'' ECOC 2018, Tu3A.4 (2018).
\bibitem{Wahls2018} S. Wahls, S. Chimmalgi and P. J. Prins, ``FNFT: A Software Library for Computing Nonlinear Fourier Transforms,'' The Journal of Open Source Software, vol. 3, no. 23, paper 597 (2018).
\end{thebibliography}
\end{document}